\documentstyle[aps,preprint,epsf]{revtex}

\begin{document}
\draft \tightenlines

\def\be{\begin{equation}}
\def\ee{\end{equation}}
\def\bea{\begin{eqnarray}}
\def\eea{\end{eqnarray}}
\def\pd{\partial}
\def\a{\alpha}
\def\b{\beta}
\def\g{\gamma}
\def\d{\delta}
\def\m{\mu}
\def\n{\nu}
\def\t{\tau}
\def\l{\lambda}

\def\s{\sigma}
\def\e{\epsilon}
\def\scri{\mathcal{J}}
\def\cM{\mathcal{M}}
\def\tcM{\tilde{\mathcal{M}}}
\def\RR{\mathbb{R}}

\preprint{KUNSAN-TP-99-6}

\title{Absorption of dilaton s-wave in type 0B string theory}

\author{ Jin Young Kim\footnote{Electronic address:
jykim@ks.kunsan.ac.kr}}
\address{Department of Physics, Kunsan National University,
Kunsan, Chonbuk 573-701, Korea}
\maketitle

\begin{abstract}
We find the absorption probability of dilaton field in type 0B
string theory.
Since the background
solutions are of the form $AdS_5 \times S^5$ on both regions,
we use the semiclassical formalism adopted in type IIB theory to find
the absorption cross section.
The background tachyon field solution was used as a reference to
relate the solutions of the two regions.
We also consider the possible corrections to
absorption probability and the $\ln(\ln z)$ form of the correction
is expected as in the calculation of the confinement solution.

\end{abstract}
\pacs{PACS number(s): 04.70.Bw, 11.27.+d}

\newpage

\section{Introduction}

Maldecena's conjecture
that ${\cal N} = 4$ super Yang-Mills theory
is dual to type IIB supergravity has opened a new horizon to the
study of large $N$ QCD\cite{MGW}. Another possibility to apply this
technique to non supersymmetric gauge theories was proposed by
Polyakov \cite{Pol9809057}. He suggested that one can study
non-SUSY gauge theories using branes in the context of type 0
string theories. The type 0 string theories are defined on the
world sheet of type II by performing a non-chiral GSO projection
\cite{DixSei}.
As a consequence type 0 theories have world sheet supersymmetry
but no space-time supersymmetry. The resulting sectors of the
theories are, in the notation of \cite{sst},
$(NS-,NS-),(NS+,NS+)$ and a doubled set of Ramond-Ramond fields.
If the doubled R-R is $(R+,R-),(R-,R+)$, the theory is type 0A and
if $(R+,R+),(R-,R-)$, the theory is 0B.
These non-SUSY string theories are also part of the web of
dualities and can be described as limits of M-theory. It is
known that the type 0A corresponds to M-theory on
${\bf S}^1 / (-1)^{F_s} \cdot S$, where $S$ denotes a
half-shift along the circle, and type 0B corresponds to M-theory
on ${\bf T}^2 / (-1)^{F_s} \cdot S$, where $S$ is along the circle
of the torus \cite{BG9906055}.

The low energy fields of the
theory are tachyon, the bosonic field of type II and two copies of
R-R fields. Accordingly they have open string
descendants which are crucial for the D-brane construction.
Following the idea of \cite{Pol9809057},
Klebanov and Tseytlin \cite{KleTse9811035} sought the possibility
that the tachyonic instability can be cured by the R-R fields.
Using the near-horizon threebrane geometry they constructed an
$SU(N)$ gauge theory with six adjoint scalars and studied its
behavior in the point of AdS/CFT correspondence.
The tachyon indeed condensed with its effective mass squared
being shifted to a positive value.
The analysis showed that the tachyon is constant at large radial
distance and the geometry is of the form $AdS_5 \times S^5$ with
vanishing coupling, which can be interpreted as a UV fixed point.
The condensation breaks the conformal invariance which agrees
with the fact that the dual gauge theory is not conformal. The
asymptotic solution for weak coupling was found by Minahan
\cite{Minahan9811156} and revealed that the logarithmic flow as
expected from the asymptotic freedom of the dual gauge theory.
Also the solution has the right sign of the $\beta$-function in
agreement with field theory and there exists
another fixed point at infinite coupling(IR fixed point),
which is also characterized
by $AdS_5 \times S^5$ Einstein frame metric.
It is argued that there is an RG trajectory that extends from
UV fixed point to the IR fixed point
\cite{KleTse9812089,Minahan9902074}.
The generic static solution to type 0 gravity equations
of motions are confining in the IR and have logarithmic scaling
in the UV.

The construction of non-SUSY branes suggests many directions of
research. And it seems that many of the results which were
obtained in type II brane configurations can be copied in the type
0 case. Since the metric solutions are asymptotically $AdS_5
\times S^5$ for both regions, the study of scattering process can
be one of the examples. Scattering amplitude or absorption cross
section involving the tachyon field cannot be addressed from the
previous work on type II theories \cite{GHKK9803023,MTR9806132}.
However one can use the same technique as in type II theories
\cite{Kleba97}
 to evaluate various scattering with D-branes in type 0
theories. The purpose of this paper is to seek this possibility.
Specifically, we will consider the semiclassical low
energy scattering of
dilaton under the static background solution of type 0B string
theory and find the absorption cross section.

The organization of the paper is as follows. In Section II we review
the type 0B theory and its background solutions of both regions
and set up some preliminaries for our calculation. In section III we
find the leading absorption cross section in the low-frequency limit.
We use the background tachyon field as a reference to relate the
solutions of both regions.
In Section IV we consider the higher order corrections
to the absorption probability.
Two kinds of orrections are possible, one from the background metric
and the other from the string coupling through the tachyon field.
Section V is devoted to conclusions.

\section{ type 0B background spacetime }
The type 0 model has a closed string tachyon, no fermions and a
doubled set of R-R fields, and thus a doubled set of D-branes
\cite{BG97}. With the doubling of R-R fields, the self-dual
constraint on the five-form field is relaxed and one can have D3
branes that are electric instead of dyonic. We consider the  low
energy world volume action of $N$ coincident electric D3 branes.
We start from the action for the type 0B
theory\cite{KleTse9812089}
\bea
S_0= & - &{ 1 \over 2 \kappa^2_{10}} \int d^{10} x \sqrt{ -g}
\bigg[ R - {1 \over 2} \nabla_n \Phi \nabla^n \Phi  \nonumber \\
     & - & {1 \over 4}  (  \nabla_n T \nabla^n T  + m^2 e^{{1 \over 2} \Phi} T^2)
 - { 1 \over 4 \cdot 5!} f(T) F_{n_1...n_5} F^{n_1...n_5} +
\cdots \bigg],
\eea
where
$g_{mn}$ is the Einstein-frame metric,
$m^2 = - {\textstyle{ 2 \over \alpha^\prime} }$, and the
tachyon-R-R field coupling function is
\be
f(T)= 1+ T + {1 \over 2}T^2.
\label{f(t)}
\ee
The equations of motion from this effective action are
\be
 \nabla^2 \Phi ={1\over 8} m^2
e^{{1\over 2}\Phi} T^2,    \label{dileq}
\ee
\bea
R_{mn} - {1 \over 2}  g_{mn} R &=& {1 \over 4} \nabla_m T \nabla_n T -
{1\over 8} g_{mn} [(\nabla T)^2 + m^2 e^{{1 \over 2}\Phi} T^2] +
{1\over 2} \nabla_m \Phi \nabla_n \Phi  \nonumber \\
&-& {1 \over 4} g_{mn} (\nabla \Phi)^2
+ {1 \over 4 \cdot 4!} f(T)
(F_{m klpq} F_{n}^{~klpq}
 - { 1\over 10} g_{mn} F_{sklpq} F^{sklpq}),   \label{graeq}
\eea
\be
 (-\nabla^2 + m^2 e^{{1\over 2}\Phi} ) T +
  {1 \over 2 \cdot 5!} f'(T) F_{sklpq} F^{sklpq} = 0,  \label{taceq}
\ee
\be
 \nabla_m [ f(T) F^{mnkpq} ] =0.  \label{RReq}
\ee
If one parametrize the 10-d Einstein-frame  metric
as in \cite{KleTse9811035}
\be
ds^2_E = e^{{1 \over2} \xi - 5 \eta} d \rho^2
+ e^{-{1 \over 2} \xi} (-dt^2 + dx_i dx_i)
 + e^{{1 \over 2}\xi - \eta} d\Omega^2_5,
\ee
where $\rho$ is the radial direction transverse to the 3-brane
($i = 1,2,3$),
then the radial effective action corresponding to
(\ref{dileq}) -- (\ref{RReq}) becomes
\be
S=  \int d \rho \bigg[{1 \over 2} \dot \Phi^2
   + {1 \over 2} \dot \xi^2
- 5 \dot \eta^2 + {1 \over 4} \dot T^2
   - V(\Phi,\xi,\eta,T) \bigg],
\ee
\be
V = {1 \over 2 \alpha^\prime} T^2
e^{{1 \over 2 }\Phi + {1 \over 2} \xi - 5 \eta }
   + 20 e^{-4\eta} - Q^2 f^{-1} (T) e^{-2\xi}.
\ee
Here the constant $Q$ is the R-R charge and dot means
the derivative with respect to $\rho$.
The resulting set of variational equations,
\be
\ddot \Phi + {1\over 4 \alpha^\prime} T^2
e^{{1 \over 2}\Phi + {1 \over 2} \xi - 5 \eta }=0, \label{Phiddot}
\ee
\be
\ddot \xi + {1\over 4 \alpha^\prime} T^2
e^{{1 \over 2}\Phi + {1 \over 2} \xi - 5 \eta } + 2 Q^2 f^{-1} (T)
e^{-2\xi}=0,   \label{xiddot}
\ee
\be
\ddot \eta + 8 e^{-4\eta } + {1\over 4 \alpha^\prime} T^2
e^{{1 \over 2}\Phi + {1 \over 2}  \xi   - 5 \eta }
=0,     \label{etaddot}
\ee
\be
\ddot T + {2 \over \alpha^\prime}
T e^{{1 \over 2}\Phi + {1 \over 2} \xi - 5 \eta }+
2 Q^2 {f^\prime (T)\over f^2(T)} e^{-2\xi}=0,  \label{Tddot}
\ee
should be supplemented by the `zero-energy' constraint
\be
{1 \over 2} \dot \Phi^2 + {1 \over 2} \dot \xi^2
- 5 \dot \eta^2 + {1 \over 4} \dot T^2
   +  V(\Phi,\xi,\eta,T) =0,  \label{zeroen}
\ee
which can be  used instead of one of the second-order equations.

In the Einstein frame the dilaton decouples from the R-R terms
($|F_5|^2$) and the tachyon mass term plays the role of the source
term.  It is easy to analyze the case where the $|F_5|^2$ is large
in (\ref{taceq}).  Assuming that the tachyon is localized near the
extremum of $f(T)$, i.e. $T=-1$, the asymptotic solution in the
UV($\rho \equiv e^{-y} \ll 1,~ y \gg 1$) region is given by
\cite{Minahan9811156,KleTse9812089}

\be
{\bar T} = -1 + {8\over y} + { 4 \over y^2}( 39 \ln y - 20)
 + O ( { \ln^2 y \over y^3}),  \label{tbsoluv}
\ee
\be
{\bar \Phi} = \ln (2^{15} Q^{-1}) - 2  \ln y  +
 {39 \over y} \ln y + O({ \ln y\over y^2 }), \label{dbsoluv}
\ee
\be
{\bar \xi} = \ln (2 Q) - y + {1\over y}
 + {1\over 2 y^2}( 39 \ln y - 104)+ O ( { \ln^2 y \over y^3}),
\label{xbsoluv}
\ee
\be
{\bar \eta} = \ln 2 - {1 \over 2}y + {1\over y}
 + {1\over 2 y^2} (39 \ln y - 38) + O ( { \ln^2 y \over y^3}).
\label{ebsoluv}
\ee
The 10-d Einstein frame metric can be written as
\bea
ds^2_E &=& R^2_0 \bigg[ ( 1 - {9\over 2y} -
{351\over 4y^2} \ln y + \cdots )   ({1\over 4}dy)^2   \nonumber  \\
  &+& (1 - { 1\over 2y} - { 39\over 4y^2} \ln y + \cdots )
  {e^{{1\over 2}y}\over 2 R^4_0} dx^\mu dx^\mu
 + (1 - { 1\over 2y} - { 39\over 4y^2} \ln y + \cdots )
 d \Omega^2_5 \bigg],    \label{gbsoluv}
\eea
where
$$ R^2_0 = 2^{-1/2} Q^{1/2} . $$
Note that with $y = 4 \ln u$, one can show that
the metric is of the form of $AdS_5 \times S^5$ at the leading
order,
\be
ds^2_E=  R^2_0  ( {du^2\over u^2} + {u^2\over 2 R^4_0} dx^\mu
dx^\mu + d \Omega^2_5  ).    \label{adsuv} \ee The corrections
cause the effective radius of $AdS_5$ to become smaller than that
of $S^5$. One can find the asymptotic freedom from the large $u$
behavior of the leading effective gravity solution. It is an
important question whether it survives the full string theoretic
treatment. It has been argued \cite{KleTse9812089} that the
solution does survive due to the special structure related to the
approximate conformal invariance.  The crucial fact is that the
Einstein metric is asymptotic to $AdS_5 \times S^5$.  This
geometry is
 conformal to flat space, so that the Weyl tensor vanishes in the
 large $u$ limit. Furthermore, both $\Phi$ and $T$ vary slowly for
 large $u$.

One interesting feature of the tachyon RG trajectory is that $T$
starts increasing from its critical value $T=-1$ from condition
$f^\prime (T) =0$. The precise form of the trajectory for finite $\rho$
is not known analytically, but the qualitative feature of the RG
equation was analyzed by Klebanov and Tseytlin
\cite{KleTse9812089,KleTse9901101}.
Since $\Phi, ~\xi$ and $\eta$ have negative second derivatives
(see equations (\ref{Phiddot}) --(\ref{etaddot})),
each of these fields may reach a maximum at some value of $\rho$.
If, for $\Phi$, this happens at finite $\rho$, then the coupling
decreases far in the infrared. As a result, a different possibility
can be realized: $\dot \Phi$ is
positive for all $\rho$, asymptotically vanishing as
$\rho \rightarrow \infty$.
They constructed the asymptotic form of such trajectory.
The point is that, as $\rho\rightarrow \infty$,
$T$ approaches zero so that, as in the UV region,
$T^2 e^{{1 \over 2} \Phi}$
becomes small. For this reason the limiting Einstein-frame
metric is again $AdS_5\times S^5$.
Thus, the theory flows to a conformally invariant
point at infinite coupling.  The explicit form of the asymptotic
large $\rho$ solution (IR solution) is
\be
{\bar T} = - {16\over y}
 - {8\over y^2} ( 9 \ln y - 3) + O ( { \ln^2 y \over y^3}),
    \label{tbsolir}
\ee
\be
{\bar \Phi} = - {1 \over 2} \ln (2 Q^2) + 2 \ln y
 - {9 \over y} \ln y + O({ \ln y\over y^2 }),
    \label{dbsolir}
\ee
\be
{\bar \xi} = {1 \over 2} \ln (2 Q^2) + y + {9\over y}
 + {9\over 2 y^2} ( 9 \ln y - {20\over 9}
) + O ( { \ln^2 y \over y^3}),  \label{xbsolir}
\ee
\be
{\bar \eta} = \ln 2 + {1 \over 2} y + {1\over y}
 + {1\over 2 y^2} ( 9 \ln y - 2) + O ( { \ln^2 y \over y^3}),
      \label{ebsolir}
\ee where $y$ is related to $\rho$ by $\rho \equiv e^{y} (\gg 1,~y
\gg 1)$. The 10-d Einstein frame metric in the IR region can be
written as \bea ds^2_E &&=  R^2_\infty \bigg[ ( 1 - {1\over 2y} -
{9\over 4y^2} \ln y + \cdots) ({1\over 4} dy)^2   \nonumber \\ &&
+ (1 - { 9\over 2y} - { 81\over 4 y^2} \ln y + \cdots )
  {e^{-{1\over 2}y}\over 2 R^4_\infty} dx^\m dx^\m
+ (1 + { 7\over 2y}  + { 63\over 4y^2} \ln y + \cdots ) d \Omega^2_5
\bigg],  \label{gbsolir}
\eea
where
$$  R^2_\infty = 2^{-3/4} Q^{1/2}. $$
The metric starts again
 as $AdS_5 \times S^5$ ($y = -4 \ln u$)
at $y =\infty$  and  becomes a negative curvature 10-d space at smaller
$y$ (bigger $u$).  As in the
large $u$  region, the  corrections
cause the effective radius of $AdS_5$ to become
smaller than that of $S^5$.

The structure of the asymptotic UV and IR solutions is
similar, suggesting that they can be smoothly connected into the full
interpolating solution.
For example, the tachyon starts at $T=-1$ at $\rho=0$, and grows
according to (\ref{tbsoluv}), then enters an oscillating regime and
finally relaxes to zero according to (\ref{tbsolir}).
Note that the coordinates corresponding to the two
regions are related by $y \to - y$.

\section{Semiclassical absorption of dilaton}

If we consider the solution found in section II as a static
background solution, one can think the low energy scattering of
string fields. Since the metric solutions are asymptotically
$AdS_5 \times S^5$ in both regions, we can use the semiclassical
formalism of type II theories\cite{Kleba97}. For simplicity of
calculation we will consider scattering of low energy dilaton
field in the type 0B string backgrounds. To study the scattering
problem, we introduce the time-dependent perturbed field of
dilaton
\be
\Phi = {\bar \Phi}(y) + \phi(y,t).
\ee
We look for the solutions to the equation of motion
for a mode of frequency $\omega$ of the dilaton
$\phi(y,t) = \phi (y) e^{i \omega t}$.
Then the linearized equation of motion governing the perturbation
of the dilaton is
\be
{1 \over \sqrt{-g}} \partial_y
\bigg( \sqrt{-g} g^{yy} \partial_y \phi(y) \bigg)
- \omega^2 g^{tt} \phi(y) + {1\over 8 \alpha^\prime}
e^{{{\bar \Phi} \over 2}} {\bar T}^2 \phi(y) = 0.
\label{phieq}
\ee

We will consider the solution in the UV region first.
To find the leading order solution in the UV region, we insert the
background metric (\ref{gbsoluv}) in the limit $y \to \infty$,
which is $AdS_5 \times S^5$, into the equation
\be
16 {d^2\phi \over dy^2} + 16 {d \phi \over dy}
      + 2 \omega^2R_0^4 e^{-{1 \over 2}y} \phi
      +{R_0^2 \over 8 \alpha^\prime}
e^{{{\bar \Phi} \over 2}} {\bar T}^2 \phi = 0,
\ee
where the background values of ${\bar T}$ and ${\bar \Phi}$ are
given by (\ref{tbsoluv}) and (\ref{dbsoluv}).
This equation becomes in terms of $u = \exp(y/4)$
\be
 {d^2\phi \over du^2} + 5 {d \phi \over du}
      + {2 \omega^2R_0^4 \over u^2} \phi
      +{R_0^2 \over 8 \alpha^\prime}
e^{{{\bar \Phi} \over 2}} {\bar T}^2 \phi = 0.
\ee
We introduce new dimensionless variable
\be
z \equiv {\sqrt2 \omega R_0^2 \over u}
= \sqrt2 \omega R_0^2 e^{-{y \over 4}}
\ee
and putting  $\phi = \psi(z)/u^2$,
we have
\be
z^2 {d^2\psi \over dz^2} + z {d \psi \over dz}
      + (z^2 -4) \psi
      +{R_0^2 \over 8 \alpha^\prime}
e^{{{\bar \Phi} \over 2}} {\bar T}^2 \psi = 0. \ee In the UV
region where the string coupling is weak, the last term is
negligible and the leading order solution is
\be
\psi(z) = H_2^{(2)}(z),
\ee
\be
\phi(z)= \bigg({z \over \sqrt2 \omega R_0^2} \bigg)^2 H_2^{(2)}(z),
\ee
where we choose the solution pure infalling at the horizon.

Now we consider the solution in the IR region.
Inserting the background metric (\ref{gbsolir}) in the
limit $y \to \infty$ into the equation, we have
\be
16 {d^2\phi \over dy^2} - 16 {d \phi \over dy}
      + 2 \omega^2R_\infty^4 e^{{1 \over 2}y} \phi
      +{R_\infty^2 \over 8 \alpha^\prime}
e^{{{\bar \Phi} \over 2}} {\bar T}^2 \phi = 0.
\ee
Here the background values of ${\bar T}$ and ${\bar \Phi}$ are
given by the solutions in the IR region (\ref{tbsolir})
and (\ref{dbsolir}).
Introducing new variable in the IR region as $v = \exp(y/4)$, the
equation becomes
\be
 {d^2\phi \over dv^2} -3 {d \phi \over dv}
      + {2 \omega^2R_\infty^4 \over v^2} \phi
      + {R_\infty^2 \over 8 \alpha^\prime}
e^{{{\bar \Phi} \over 2}} {\bar T}^2 \phi = 0.
\ee
Note that $u$ and $v$ are related by inverse because $y$'s in the UV
and IR regions are related by $y \to - y$.
In terms of new dimensionless variable
\be
\sigma \equiv \sqrt2 \omega R_\infty^2 v = \sqrt2 \omega
R_\infty^2 e^{{y \over 4}} \ee and with $\phi = v^2 f(\sigma)$, we
have
\be
\sigma^2 {d^2f \over d \sigma^2} + \sigma {d f \over d \sigma}
      + (\sigma^2 -4) f
      +{R_\infty^2 \over 8 \alpha^\prime}
e^{{{\bar \Phi} \over 2}} {\bar T}^2 f = 0.
\ee
In the IR region, we can also neglect the last term since
${\bar T} \to 0$ in the leading order and the solution is
given by the Bessel functions
\be
f(\sigma) = \alpha J_2(\sigma) + \beta J_{-2}(\sigma),
\ee
where the first term corresponds to the normalizable mode and the
second to nonnormalizable mode.
We choose $\beta = 0$ for the solution to be finite as
$\sigma \to \infty$, then
\be
\phi(\sigma)= \alpha \bigg({\sigma \over \sqrt2 \omega R_\infty^2} \bigg)^2
       J_2 (\sigma).
\ee

To relate the solution in the UV region and IR region,
 we adopt the matching scheme similar to the cases of D3-brane
and M5-branes \cite{GHKK9803023} and D1-D5 brane system
\cite{MTR9806132}. For both cases the equation of motion for the
minimal scalar has a self-dual point defined by the radius of the
effective anti-de Sitter space. In our case we have to be careful
in choosing the matching point because the effective radii of the
anti-de Sitter space for both regions are different because of the
tachyon coupling. However, the tachyonic coupling is always
combined with the background dilaton $\bar \Phi$, i.e. the
coupling is of the form of $e^{{\bar \Phi}/2} {\bar T^2}$. And it
seems that the functional behavior of the coupling terms for both
UV and IR region are the same. To see this, we calculate the
coupling terms up $O(1/y)$. Substituting the UV background
solution (\ref{tbsoluv}) and (\ref{dbsoluv}) into the coupling
term, we have
\be
  R_0^2 e^{{{\bar \Phi} \over 2}} {\bar T}^2 =
  2^{-{1 \over 2}} Q^{{1 \over 2}} \cdot 2^{{15 \over 2}}
 Q^{-{1 \over 2}} {1 \over y} \cdot 1^2 = {2^7 \over y}.
\ee
Similarly, from (\ref{tbsolir}) and (\ref{dbsolir}), we have
\be
  R_\infty^2 e^{{{\bar \Phi} \over 2}} {\bar T}^2 =
  2^{-{3 \over 4}} Q^{{1 \over 2}} \cdot 2^{-{1 \over 4}}
 Q^{-{1 \over 2}} y \cdot {16 \over y}^2 = {2^7 \over y}.
\ee
 Up to the order of $O(1/y)$, the functional form of the
perturbed terms are the same. And this may be true for higher
orders too. This implies that the functional forms of the
correction will be the same. Thus if we extend the UV solution
toward the IR region and IR solution toward the UV region, they
can be matched at some point of $y$.  But we don't know the exact
position of $y$. So we choose the matching point from the point
where the background tachyon fields have the same values at the
leading order.
 Equating (\ref{tbsoluv}) and (\ref{tbsolir}) in
the leading order, $-1 +8/y = -16/y$, we have $y=24$ as
the matching point.  In terms $\rho$, this corresponds that
we match the asymptotic solution of the UV region,
which is valid for very small $\rho$, at $\rho =e^{-24}$ and
that of the IR region, valid for very large $\rho$, at
$\rho = e^{24}$.
Matching the amplitude of $\phi$ at the leading
order
\be
\bigg({z \over \sqrt2 \omega R_0^2} \bigg)^2 H_2^{(2)}(z)
       \bigg |_{z=\sqrt2 \omega R_0^2e^{-6} }
= \alpha \bigg({\sigma \over \sqrt2 \omega R_\infty^2} \bigg)^2
       J_2 (\sigma)
        \bigg |_{\sigma=\sqrt2 \omega R_\infty^2e^{-6} },
\ee
one finds the relative coefficient relating the solutions of two
regions.
\be
\alpha = {8 i \over \pi}{e^{24} \over \omega^4 R_0^4 R_\infty^4}
\propto
{1 \over \omega^4 Q^2}.
\ee

The asymptotic form of the infalling wave function in the UV region is
\be
\phi(z) =
\bigg({z \over \sqrt2 \omega R_0^2} \bigg)^2 H_2^{(2)}(z)
\simeq {1 \over (\sqrt2 \omega R_0^2)^2 } \sqrt{2 \over \pi}
  z^{3 \over 2} \exp \{ -i(z - {3 \over 4} \pi) \}.
\ee
From this the invariant flux, defined by
\be
F_{\rho \to 0} = {1 \over 2i}
( \phi^* \sqrt{-g} g^{yy} \partial_y \phi -
  \phi \sqrt{-g} g^{yy} \partial_y \phi^*) \bigg |_{y=\infty},
\ee
is given by
\be
F_{\rho \to 0} = {1 \over 2 \pi}.
\ee
Since the ingoing part of the wavefunction in the IR region is
\be
\phi(\sigma) = \alpha
\bigg({\sigma \over \sqrt2 \omega R_\infty^2} \bigg)^2 J_2(\sigma)
\simeq {1 \over (\sqrt2 \omega R_\infty^2)^2 } \sqrt{1 \over 2 \pi}
    \sigma^{3 \over 2}\exp \{ i(\sigma - {3 \over 4} \pi ) \},
\ee
the ingoing flux at infinity is given by
\be
F_{\rho \to \infty} = {|\alpha|^2 \over 8 \pi}
= {8 e^{48} \over \pi^3 \omega^8 R_0^8 R_\infty^8}.
\ee
The absorption probability is given by the ratio of the
flux in the UV region to the ingoing flux in the IR region and we
find
\be
P = {F_{\rho \to 0} \over F_{\rho \to \infty}} = {\pi^2 \omega^8
R_0^8 R_\infty^8 \over 16 e^{48}} \propto \omega^8 Q^4. \ee The
s-wave absorption cross-section is given by\cite{Kleba97}
\be
\sigma^s_{abs} = { (2 \pi)^{d-1} \over \omega^{d-1} \Omega_{d-1} }
P \bigg|_{d=6} ={ \pi^4 \over  e^{48} } R_0^8 R_\infty^8 \omega^3
\propto \omega^3 Q^4. \ee
 Note that the cross section has an
$\omega^3$ dependence, similar to the type II case. If one chooses
different position of $y$ as the matching point, the numerical
coefficient can be changed but the $\omega$ and $Q$ dependence
will be the same.

\section{Higher order corrections}

In this section we will consider the effect of higher order
corrections of the dilaton
absorption cross-section. The dominant corrections to absorption
cross-section arise from the matching about the point where the
tachyon field have the same value. In the case of D3-brane and
M5-brane of type IIB theory the correction to the minimally
coupled scalar equation is done by considering the higher order
terms of the background metric. However, in our case where the
dilaton is coupled to the tachyon through the string
coupling $\alpha^\prime$, we have to consider the correction
by this coupling term too. For simplicity of the calculation we
consider only the leading terms and subleading corrections
will be straightforward.

In the UV region, inserting
$$
\sqrt{-g} = {R_0^2 e^y \over 16} (1 -{9 \over 2y}+ \cdots), ~
g^{yy} = {16 \over R_0^2} (1 +{9 \over 2y}+ \cdots),~
g^{tt} = - 2 R_0^2 e^{-{1 \over 2} y} ( 1 + {1 \over 2 y} +
\cdots)
$$
and
$$
\bar T^2 = 1 - {16 \over y} + \cdots,~~
e^{\bar \Phi \over 2} = {2^7 \over R_0^2}
( {1 \over y} + \cdots )
$$
into (\ref{phieq}), we have
\be
16 {d^2\phi \over dy^2} + 16 {d \phi \over dy}
+ 2 \omega^2R_0^4 e^{-{1 \over 2}y} ( 1 + { 1 \over 2 y} + \dots) \phi
   + {16 \over \alpha^\prime} ( { 1 \over y} + \cdots) \phi = 0.
\ee
Substituting $ y = 4 \ln u$ and $\phi = \psi/u^2$, this can be written as
\be
 {d^2 \psi \over du^2} + {d \psi \over du}
      + ( {2 \omega^2R_0^4 \over u^2} - 4) \psi
 = \bigg\{ {2 \omega^2R_0^2 \over u^2} ( { 4 \over y} + \cdots)
      -{16 \over \alpha^\prime} ( {1 \over y} + \cdots) \bigg\}\psi.
\ee
With the dimensionless variable defined earlier
$z = {\sqrt2 \omega R_0^2 \over u}
= \sqrt2 \omega R_0^2 e^{-{y \over 4}}, $
we get
\be
z^2 {d^2\psi \over dz^2} + z {d \psi \over dz}
      + (z^2 -4) \psi =
      \bigg\{ {z^2 \over \ln {\sqrt2 \omega R_0^2 \over z} }
      (1 + \cdots)
      - {4 \over \alpha^\prime}
      {1 \over \ln({\sqrt2 \omega R_0^2 \over z})}
      (1 + \cdots) \bigg\}  \psi.
\ee
In the low energy scattering case the terms on the right hand side act
as small perturbations. The first term on the right hand side is
due to metric correction while the second is from the tachyon coupling.
We look for the perturbative solution of the form
 \cite{GHKK9803023,MTR9806132,KLM9812016}
\be
\psi(z) = \psi_0(z) + \psi_1(z) + \cdots,
\ee
where the zeroth order solution is
\be
\psi_0(z) = H_2^{(2)} (z).
\ee
$\psi_1$ satisfies the inhomogeneous equation
\be
z^2 {d^2\psi_1 \over dz^2} + z {d \psi_1 \over dz}
  + (z^2 -4) \psi_1 =
 \bigg\{  {z^2 \over \ln {\sqrt2 \omega R_0^2 \over z} }(1 +\cdots)
  - {4 \over \alpha^\prime}
  {1 \over \ln {\sqrt2 \omega R_0^2 \over z} } (1 +\cdots) \bigg\} \psi_0.
\ee
The solution to this equation is
\be
\psi_1(z) = {\pi \over 2} \int^z d x {1 \over x} \bigg(
{- x^2 \over \ln {\sqrt2 \omega R_0^2 \over x} }
      + {4 \over \alpha^\prime}
      {1 \over \ln {\sqrt2 \omega R_0^2 \over x}} \bigg)
      H_2^{(2)}(x) \{ J_2(x) Y_2(z) - J_2(z) Y_2(x) \}.
\label{psi1cor}
\ee
Substituting the series form of Bessel functions,
and retaining the terms necessary for given order,
one can find the correction in series form.

We repeat the same procedure in the IR region to find the
higher order solution of the form
\be
f(\sigma) = f_0(\sigma) + f_1(\sigma) + \cdots,
\ee
where $f_0(\sigma)$ is the zeroth order solution given by
$f_0(\sigma) = \alpha J_2(\sigma).$
The inhomogeneous equation which $f_1$ should satisfy is
\be
\sigma^2 {d^2f_1 \over d\sigma^2} + \sigma {d f_1 \over d \sigma}
      + (\sigma^2 -4) f_1 =
 -\bigg\{ { \sigma^2 \over \ln { \sigma \over \sqrt2 \omega R_\infty^2 } }
 (1 +\cdots)
     +  {4 \over \alpha^\prime}
      {1 \over \ln {\sigma \over \sqrt2 \omega R_\infty^2 } }
      (1 +\cdots) \bigg \} f_0,
\ee
and the solution is
\be
f_1(\sigma) = \alpha {\pi \over 2} \int^\sigma d x {1 \over x}
\bigg \{
{x^2 \over \ln {x \over \sqrt2 \omega R_\infty^2 } } (1 + \cdots)
      +  {4 \over \alpha^\prime}
      {1 \over \ln {x \over \sqrt2 \omega R_0^2}} (1 + \cdots) \bigg \}
      J_2(x) \{ J_2(x) Y_2(\sigma) - J_2(\sigma) Y_2(x) \}.
\label{f1cor}
\ee
Matching the two solutions at the matching point
\be
\bigg({z \over \sqrt2 \omega R_0^2} \bigg)^2
( \psi_0(z) + \psi_1(z) )
       \bigg |_{z=\sqrt2 \omega R_0^2e^{-6} }
= \alpha \bigg({\sigma \over \sqrt2 \omega R_\infty^2} \bigg)^2
       f_0(\sigma) + f_1(\sigma)
        \bigg |_{\sigma=\sqrt2 \omega R_\infty^2e^{-6} },
\ee
one can find the correction of the relative coefficient $\alpha$.

If we substitute the series expansion of Bessel functions,
it is expected the that the leading correction is
of the form $\ln(\ln z)$ from the integral structure
$$\int^z {dx \over x} {x^n \over \ln x} = z^n \ln (\ln z) + \cdots.$$
However, if we retain only the leading terms of the  the series
expansion of Bessel functions
\be
J_2(x) = {x^2 \over 8},~~Y_2(x)= - {4 \over \pi x^2},~~
H_2^{(2)}(x) ={4 i \over \pi z^2},
\ee
the correction terms from this type cancels out. The correction
of the relative coefficient $\alpha$
is from the subleading terms i.e. next order terms of Bessel functions
and $\cdots$ of (\ref{psi1cor}) and (\ref{f1cor}).
This is a good consistency check for the background solutions.

\section{Discussion}

We considered the low energy scattering of dilaton field in the
type 0B theory.
Using the fact that the asymptotic structure of the background
solutions on both regions is $AdS_5 \times S^5$, we calculated
semiclassically the absorption probability.
Since the tachyonic coupling term is small on both
regions, the calculation is similar to the case of minimally
coupled scalar of type II theory.
A crucial role in our formalism was played by the background
tachyon field. We used it as a reference to find the matching
point. We also considered the possible corrections of
absorption probability and the $\ln(\ln z)$ form of the correction
\cite{Minahan9811156,KleTse9812089}
is expected as in the calculation of the confinement solution.

In the present paper, we perturbed only the dilaton field from the
background for simplicity of calculation. The
right way to analyze the scattering is to perturb all fields in the
theory and solve the full coupled equations, which seems very
difficult to solve.
We considered the low energy world volume
action of $N$ coincident electric D3 branes. We expect that the
same formalism can be applied to self-dual 3-branes
\cite{KleTse9901101} because the
asymptotic structure of the background also has $AdS_5 \times S^5$
structure. Though we considered only the case of $d=10$ critical case,
it seems interesting to study the case of non critical type 0 theory in
lower dimensions \cite{noncri}.

\section*{Acknowledgement}
I would like to thank S.P.Kim, Y.S.Myung and H.W.Lee for useful
discussions. This work was supported by the Korean Physical
Society (1999).

\end{document}